\documentclass[12pt]{iopart}
\usepackage{epsfig}
\usepackage{graphicx}

\begin{document}

\title{Roughness and multiscaling of planar crack fronts}

\author{Lasse Laurson$^1$ and Stefano Zapperi$^{1,2}$}

\address{$^1$ISI Foundation, Viale S. Severo 65, 10133 Torino, Italy}
\address{$^2$IENI-CNR, Via R. Cozzi 53, 20125, Milano, Italy}

\eads{\mailto{lasse.laurson@gmail.com}, \mailto{stefano.zapperi@cnr.it}}

\begin{abstract}
We consider numerically the roughness of a planar crack front within the
long-range elastic string model, with a tunable disorder correlation length
$\xi$. The problem is shown to have two important length scales, $\xi$ and the
Larkin length $L_c$. Multiscaling of the crack front is 
observed for scales below $\xi$, provided that the disorder is strong enough. 
The asymptotic scaling with a roughness exponent $\zeta \approx 0.39$ 
is recovered for scales larger than both $\xi$ and $L_c$. If $L_c > \xi$,
these regimes are separated by a third regime characterized by the Larkin
exponent $\zeta_L \approx 0.5$. We discuss the experimental implications of 
our results.
\end{abstract}


\maketitle

\section{Introduction}\label{intro}
Crack fronts propagating along a weak plane of a disordered material
provide an ideal experimental system for studying phenomena such as
the depinning transition, the associated avalanche dynamics and 
roughness of the crack front. Experimentally, by studying systems
such as sintered Plexiglas plates \cite{SCH-97,DEL-99,MAL-06} and paper 
sheets \cite{SAL-06}, it has been established
that a slowly driven planar crack front propagates in a sequence of avalanches,
with a power law size distribution. In the case of the Plexiglas experiments
the crack fronts have been observed to be rough \cite{SCH-97,DEL-99}. 

As is often the case when studying similar phenomena typically exhibiting some degree of 
universality, it is expected that the statistical properties of the 
planar crack fronts can be theoretically described with a simplified model, 
which here is given by the long-range elastic string \cite{GAO-89,SCH-95}. 
Such a model has successfully reproduced the statistics of avalanches of 
crack front propagation \cite{BON-08}, and in particular the local clusters 
such avalanches are broken into due to the long range elastic interactions 
between different segments of the crack front \cite{LAU-10}. Early experiments 
\cite{SCH-97,DEL-99} found a roughness exponent significantly larger than 
that predicted by the crack line model \cite{SCH-95,ROS-02,DUE-07}, but a more recent 
study has shown that the theoretically predicted value of the roughness 
exponent is recovered if large enough length scales are considered 
\cite{SAN-10}. Moreover, the short length scale regime has been found to 
exhibit multiscaling \cite{SAN-10}.

In this paper we consider the long range elastic string driven in a 
random potential with a tunable disorder correlation length, in order to 
clarify the origin and nature of the different scaling regimes of the 
crack front roughness. The main benefit of our model as compared to 
previous studies of the crack line model is that the disorder correlation
length $\xi$ can be chosen to be larger than the crack line segment length,
thus making it possible to study also the regime below $\xi$. It is 
demonstrated that the problem includes two 
relevant length scales, the disorder correlation length $\xi$ and the 
Larkin length $L_c$ \cite{LAR-79}. By considering different sets of parameters of the 
model, the relative magnitudes of these two length scales can be tuned. 
For strong enough disorder (or equivalently soft enough lines), 
multiscaling of the crack front is observed for scales below $\xi$. 
Asymptotically, for length scales exceeding both $\xi$ and $L_c$, we 
recover the well known roughness exponent $\zeta \approx 0.39$ of the 
long-range elastic string \cite{SCH-95,ROS-02,DUE-07}. An intermediate regime 
characterized by the Larkin exponent $\zeta_L \approx 0.5$ is observed above 
$\xi$ and below $L_c$, if $\xi<L_c$. The paper is organized as follows: In the next
Section, the details of the numerical model are presented. In Section \ref{sec3}
we present a theoretical description of the expected scaling regimes
of the front roughness, while Section \ref{sec4} includes 
detailed numerical results from simulations of the crack line model.
Section \ref{sec5} finishes the paper with conclusions and discussion.

\section{Model}

\begin{figure}[!ht]
\begin{center}
\includegraphics[scale = 0.66667,clip]{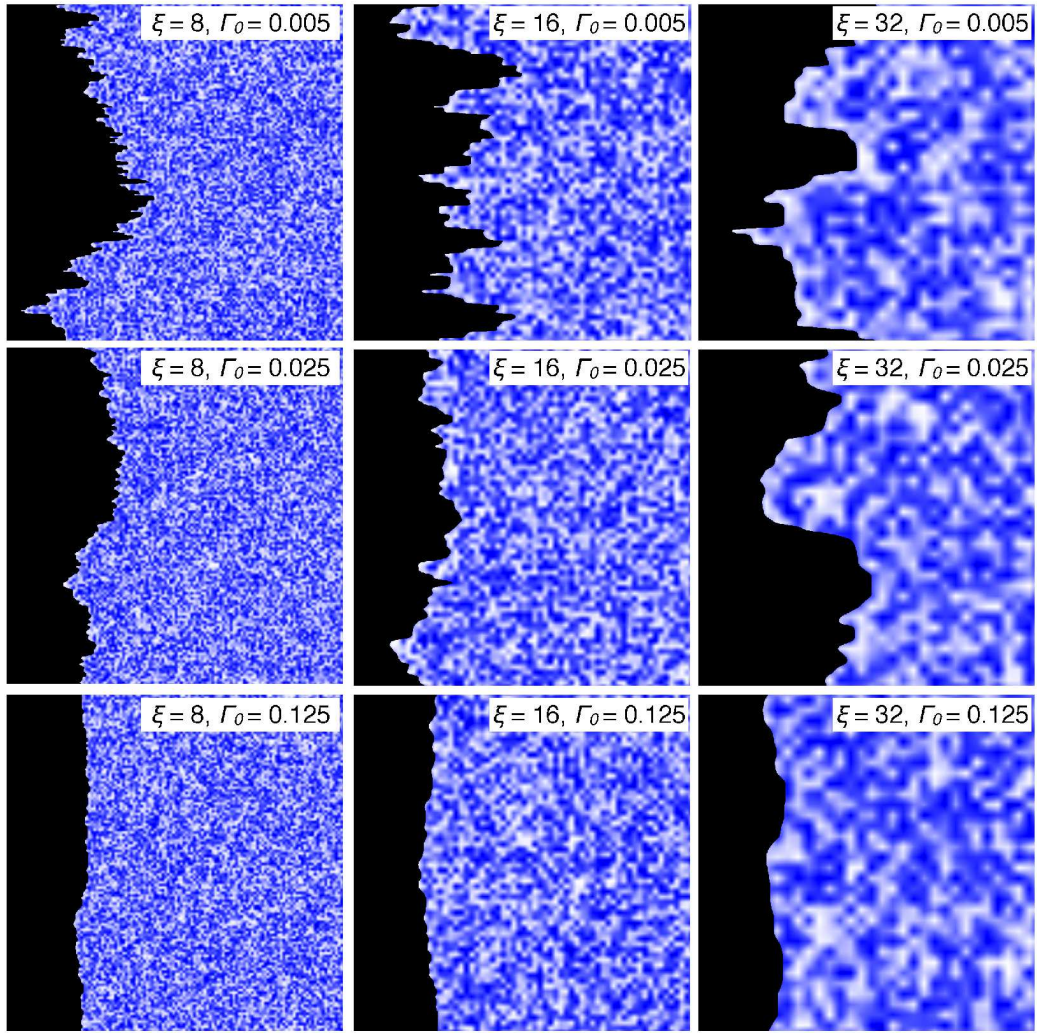}
\end{center}
\caption{Snapshots of crack fronts from the line model moving from left to
right, with various
values of $\xi$ and $\Gamma_0$. For each case, a square area of the
weak plane of linear size $L=1024$ is shown. The cracked region is displayed in
black. The different shades of blue indicate different
values of the local toughness of the non-cracked material: dark (light)
blue areas correspond to weak (strong) material.}
\label{fig:fronts}
\end{figure}

The model of a propagating planar crack front considered here is represented
by a vector of single-valued integer heights $h_i$, $i=1,\dots,L$, with $L$ the system
size \cite{SCH-95}. Crack propagation is driven by the local stress intensity factor
(SIF) $K_i$, representing the asymptotic prefactor of the $1/\sqrt{r}$
divergence if the stress field near the crack tip. $K_i$ is taken to
be of the form $K_i = K_i^{elastic} + K_{i,h_i}^{random}+K^{ext}$. Here
\begin{equation}
K_i^{elastic} = \Gamma_0 \sum_{j \neq i}^L \frac{h_j - h_i}{b|j-i|^2}
\end{equation}
is the first-order variation of the stress intensity factor due to first-order
perturbation of the front position \cite{GAO-89}, with $b$ the front segment spacing
and $\Gamma_0$ tunes the strength of the long-range elastic interactions. 
For periodic boundary conditions along the crack front as considered here, 
this becomes \cite{TAN-98,ROS-08}
\begin{equation}
K_i^{elastic} = \Gamma_0 b \left(\frac{\pi}{L}\right)^2
\sum_{j \neq i}^L \frac{h_j - h_i}{\sin^2(|j-i| b \pi/L)}.
\end{equation}
$K_{i,h_i}^{random}$ is a time-independent disorder field with
correlations
\begin{equation}
\langle K_{i,h_i}^{random} K_{j,h_j}^{random} \rangle 
- \langle K_{i,h_i}^{random} \rangle  \langle K_{j,h_j}^{random} \rangle \sim  f(r/\xi),
\end{equation}
where $r=\sqrt{ (i-j)^2 + (h_i-h_j)^2 }$ and $f(x)=1$ for $x\ll1$ and
$f(x)=0$ for $x \gg 1$. In practice, such a disorder
field can be prepared by assigning uncorrelated random numbers from
a distribution to a subset of lattice sites forming a square grid with
a spacing of $\xi$, and filling the rest of the lattice sites by
using an appropriate interpolation algorithm. Here, we use a uniform
distribution from -1 to 0, and bilinear interpolation to obtain
a smooth disorder field, see Fig. \ref{fig:fronts} for examples. $K^{ext}$ is the 
contribution of the external load. The dynamics is defined in discrete 
time by setting
\begin{equation}
v_i(t)=h_i(t+1)-h_i(t) = \theta(K_i),
\end{equation}
where $v_i$ is the local velocity of the front element $i$, and $\theta$
is the Heaviside step function. Parallel dynamics is assumed, such that
during a single time step, all the front elements with $v_i>0$ are
advanced by a unit step, $h_i(t+1)=h_i(t)+1$. The local forces are then
recomputed for each element, and the process is repeated until $v_i=0$
for all $i$ and the avalanche stops. The applied load is then increased
so that exactly one front element becomes unstable (i.e. $v_i>0$), and
a new avalanche is initiated. As the crack front advances, the applied
SIF $K^{ext}$ decreases at a rate proportional to the instantaneous
average velocity $v(t)=1/L \sum_{i=1}^L v_i(t)$ of the front, with a 
proportionality constant $\epsilon$. Such a protocol leads to a cut-off for the
avalanche size distribution scaling with $\epsilon$ \cite{LAU-10}, and corresponds 
to quasistatic driving, which has the advantage that observables such as 
the avalanche sizes can be defined without the need to apply (a possibly 
ambiguous) non-zero threshold. Notice that while the discretized
dynamics we employ here neglects the fact that the local velocity of
the crack front should be proportional to the local SIF, this simplification
is well known to have no influence on the scaling behaviour we study here.

\section{Roughness of the crack front due to disorder}\label{sec3}

As the planar crack front advances along the weak plane, it will roughen
due to disorder. Such disorder, originating from the fluctuations of the local 
toughness of the weak plane, is characterized by a fluctuation amplitude 
$R$ and a correlation length $\xi$. The crack front morphology
can be characterized by considering the structure functions \cite{SAN-10,SAN-07}
\begin{equation}
C_k(\delta) = \langle |h_{i+\delta}-h_i|^k \rangle_i^{1/k}.
\end{equation}
If the front fluctuations follow Gaussian statistics, the structure
functions with different $k$ can be collapsed by normalizing the $C_k$'s
with the Gaussian factors \cite{SAN-10,SAN-07}
\begin{equation}
R_k^G = \sqrt{2} \left[ \frac{\Gamma ((k+1)/2)}{\sqrt{\pi}} \right]^{1/k}.
\end{equation}
Close to the depinning transition (i.e. when a slow enough external 
drive is applied), we expect the roughness of
the front to be controlled by two length scales, $\xi$ and the Larkin length 
$L_c \sim \Gamma_0^2 \xi / R^2$ \cite{LAU-10,MOR-04}. In the following, we will 
consider the different regimes separately.

\subsection{Strong pinning: multiscaling of the front}

For small enough $\Gamma_0$ (or equivalently, large enough $R$), the
disorder is strong enough to substantially deform the front locally,
which may lead to steep slopes (and possibly even overhangs in experiments
and in models which would allow them to be formed) of the crack front, 
and thus to multiscaling of the front roughness, or $k$-dependent 
scaling of the structure functions $C_k(\delta)$. In the spirit of the
idea of strong pinning (individual ``pinning centers'', or correlated
fluctuations of the random potential,  are strong enough to substantially 
deform the front), we expect such large deformations and steep slopes 
of the crack front to take place up to a lateral length 
scale proportional to the disorder scale $\xi$. This regime corresponds to 
a small Larkin length as compared to the scale of the toughness fluctuations, 
$L_c \ll \xi$. Thus, we expect to observe multiscaling for scales
below $\xi$, and the asymptotic scaling characterized by the unique 
roughness exponent $\zeta \approx 0.39$ for scales larger than $\xi$. 
By tuning $\Gamma_0$ within this regime, the amplitude of the local
deformations and thus the strength of the multiscaling can be 
changed.

\subsection{Weak pinning: the Larkin regime}

In the opposite limit, where $L_c \gg \xi$, the toughness fluctuations
are weak and no significant deformation of the crack front takes place 
for lateral scales of the order of $\xi$. Instead, the front remains
essentially undeformed for lateral scales below $L_c$, and the potential
energy fluctuations follow Poissonian statistics. This leads to front
roughness characterized by the Larkin exponent $\zeta_L=1/2$ for scales
above $\xi$ and below $L_c$. For scales above $L_c$ the toughness 
fluctuations become effective and the asymptotic roughness exponent 
$\zeta \approx 0.39$ is expected.

\begin{figure}[!ht]
\begin{center}
\includegraphics[width = 5.7cm,angle=-90,clip]{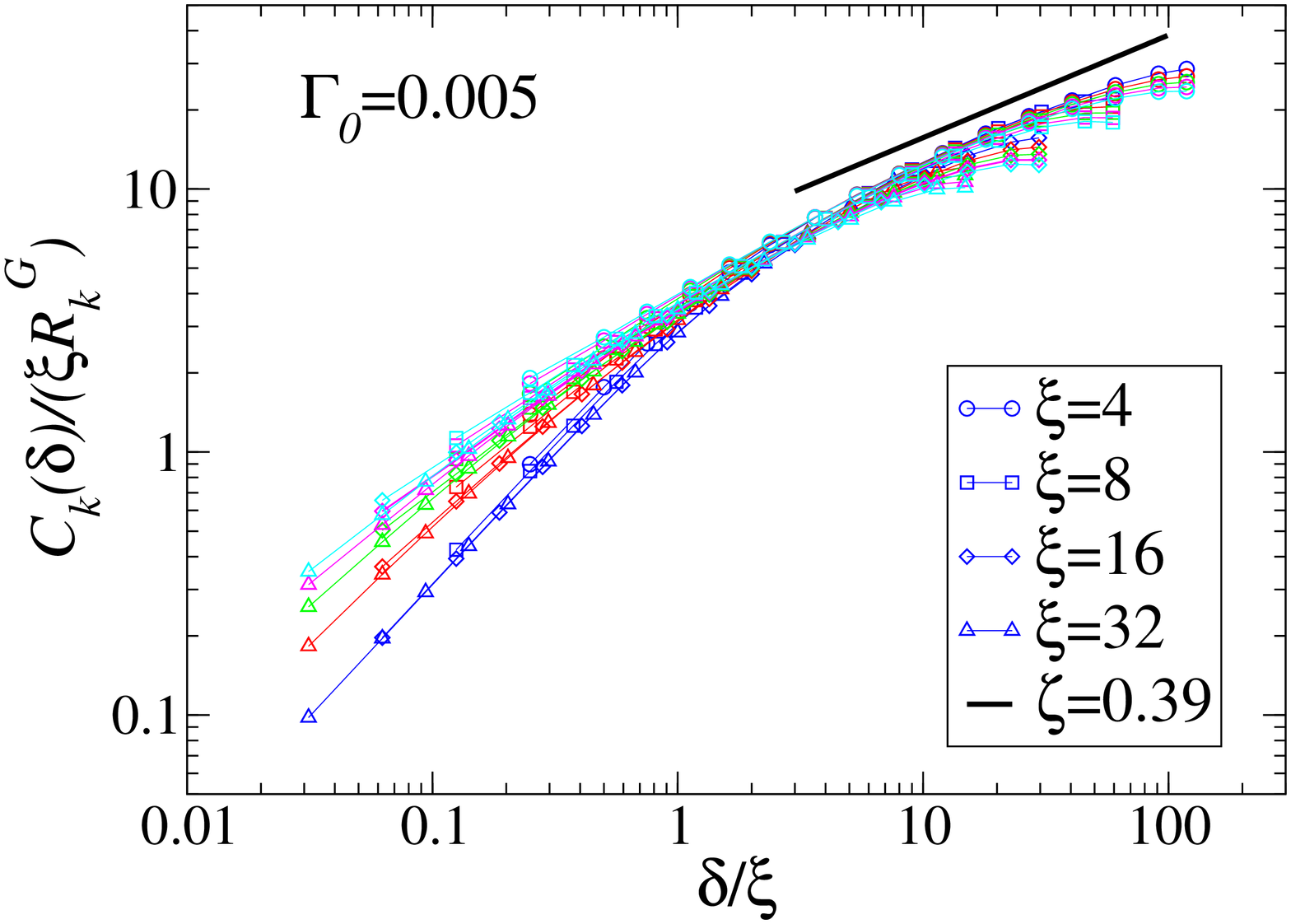}\\
\vspace{0.3cm}
\includegraphics[width = 5.7cm,angle=-90,clip]{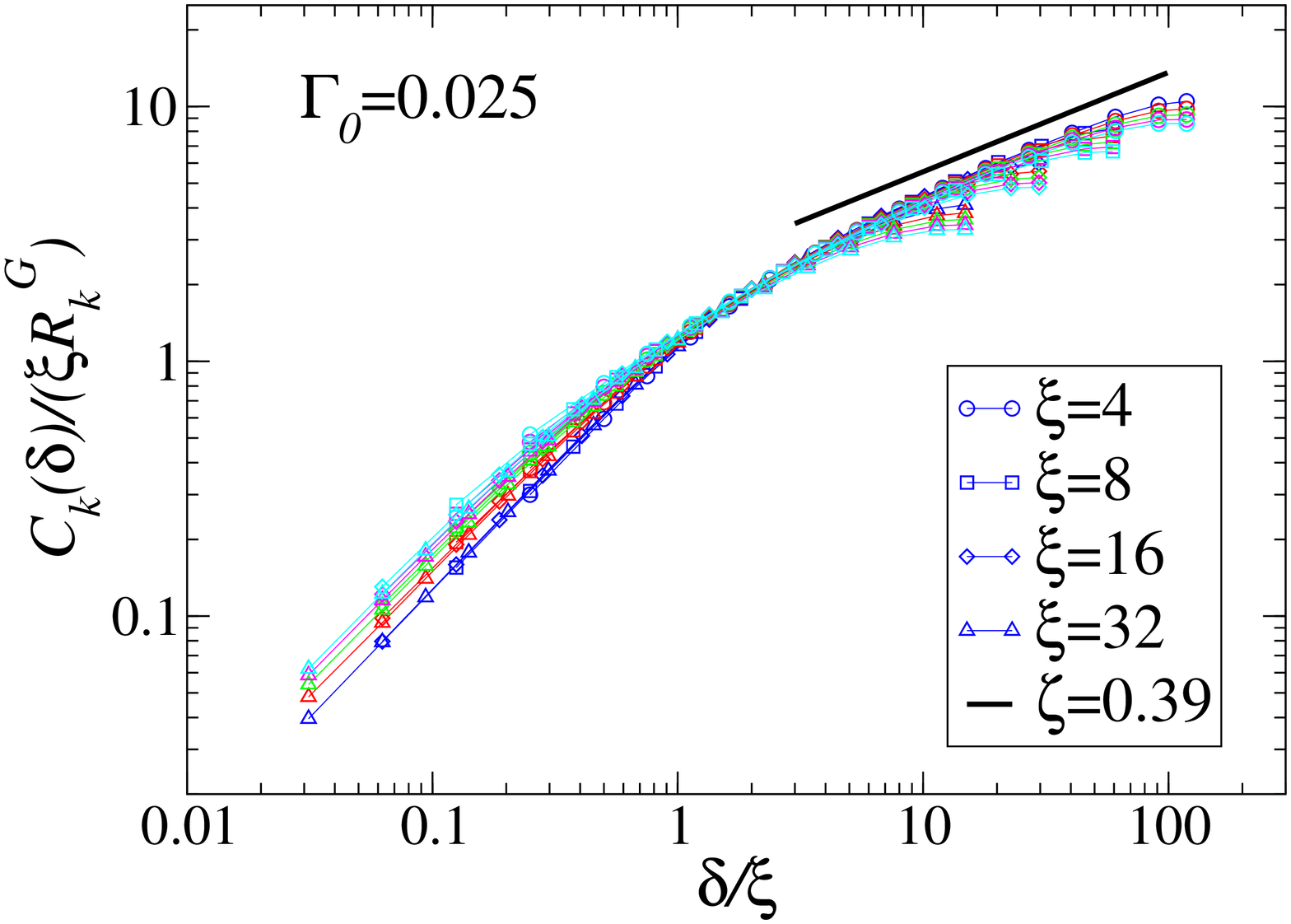}\\
\vspace{0.3cm}
\includegraphics[width = 5.7cm,angle=-90,clip]{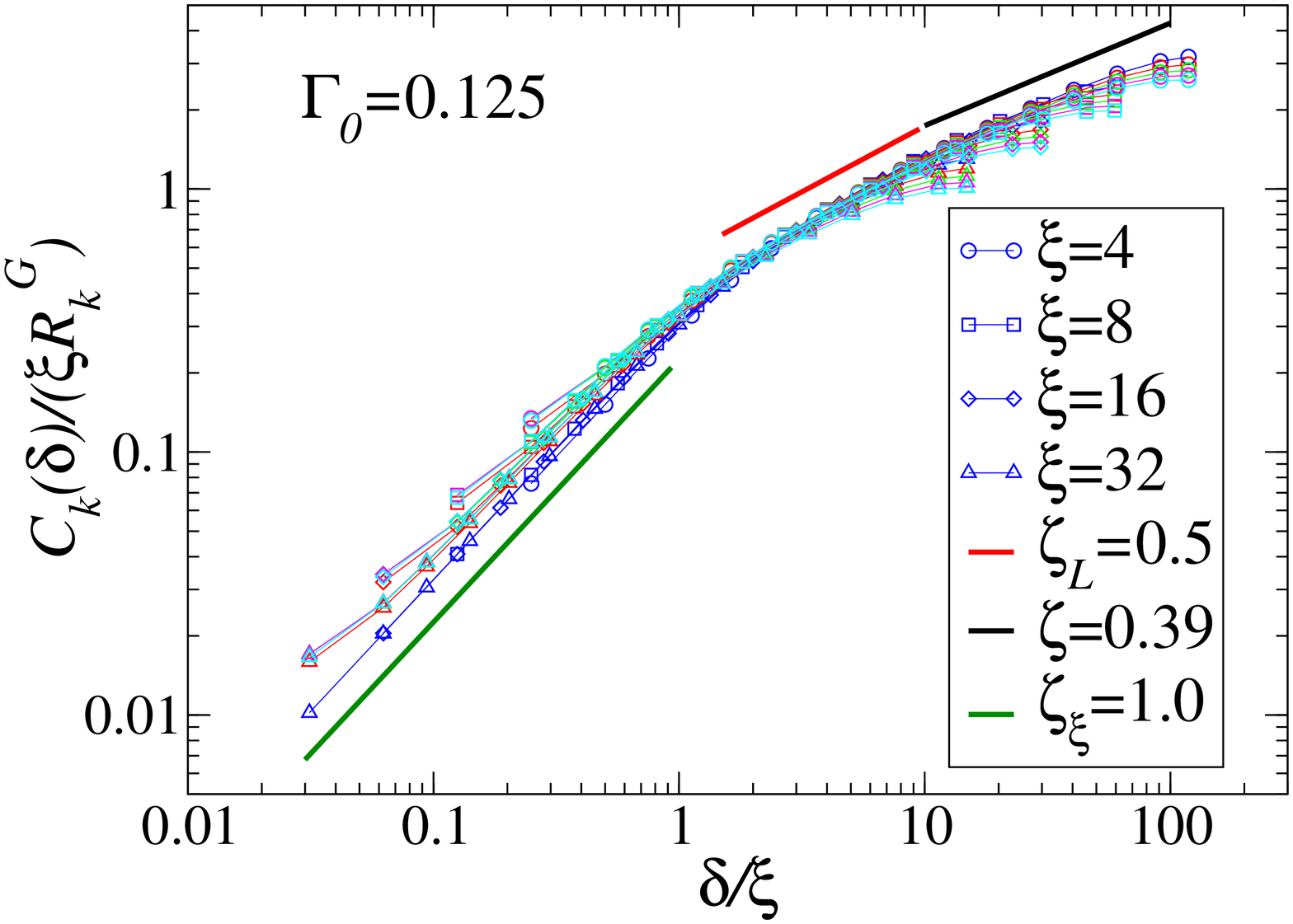}
\end{center}
\caption{The structure  functions $C_k(\delta)$ rescaled according to
Eq. (\ref{eq:collapse}). Various
values of $\xi$ (different symbols, as indicated by the legends) and $k$ (different
colors; blue, red, green, magenta and cyan correspond to $k=1$, 2, 3, 4
and 5, respectively) are considered. Data for three values of the stiffness
parameter $\Gamma_0$ are shown: $\Gamma_0=0.005$ (top), $\Gamma_0=0.025$ (middle),
and $\Gamma_0=0.125$ (bottom).}
\label{fig:roughness}
\end{figure}

\section{Numerical results}\label{sec4}

We simulate the crack line model in a system of linear size $L=1024$,
by considering different disorder correlation lengths ($\xi=4, 8, 16$ 
and $32$) and values of the crack front stiffness ($\Gamma_0 = 0.005, 
0.025$ and $0.125$). The value $\epsilon=0.001$ is used for the 
parameter controlling the avalanche cut-off scale. Fig. \ref{fig:fronts} shows
examples of the crack front profiles for different $\xi$ and $\Gamma_0$.
Also the toughness fluctuations are shown, with dark (light) blue
corresponding to regions with low (high) local toughness.

In the spirit of the discussion of the previous Section, we consider 
the scaling form
\begin{equation}
\label{eq:collapse}
C_k(\delta) = R_k^G \xi F(\delta/\xi)
\end{equation}
to account for the effect of varying $\xi$ for fixed $\Gamma_0$.
Here, $F(x)$ is expected to scale as 
\begin{equation}
\label{eq:fx11}
F(x) \sim x^{\zeta}
\end{equation} 
for $x>1$ and $L_c \ll \xi$, and 
\begin{eqnarray}
\label{eq:fx12}
F(x) & \sim x^{\zeta_L} \ \ & \textrm{for} \ \ 1<x<L_c/\xi, \\
\label{eq:fx22}
F(x) & \sim x^{\zeta} \ \ \ & \textrm{for} \ \ x>L_c/\xi
\end{eqnarray}
for $L_c \gg \xi$. Fig. \ref{fig:roughness} shows the rescaled 
$C_k(\delta)$ functions for three different values of $\Gamma_0$. 
For $\delta/\xi>1$, the data collapse nicely, verifying the scaling 
form, Eq. (\ref{eq:collapse}), and the Gaussian nature of the large 
scale front roughness. Also the different scaling regimes as given by 
Eqs. (\ref{eq:fx11}), (\ref{eq:fx12}) and (\ref{eq:fx22}) are 
clearly visible, with $\zeta \approx 0.39$ and $\zeta_L \approx 0.5$. 
Notice also that no multiscaling is observed, as long as only scales 
above $\xi$ are considered.

For $\delta/\xi<1$ and small $\Gamma_0$, the data displays multiscaling,
a signature of a deviation from the Gaussian statistics. The strength of 
this multiscaling depends on $\Gamma_0$.  This indicates
that the multiscaling regime may not be universal: different parameters
of the model could correspond to different effective exponents within 
this regime. For large enough $\Gamma_0$, the $\delta/\xi<1$ regime is
characterized by a linear dependence of $C_k(\delta)$ on $\delta$, or
$C_k(\delta) \sim \delta^{\zeta_{\xi}}$ with $\zeta_{\xi} \approx 1.0$, 
and essentially no multiscaling is observed: The small deviation from
linear behaviour for the smallest values of $\delta$ in the bottom panel 
of Fig. \ref{fig:roughness} is due to the discrete nature of the model.

\section{Conclusions and discussion}\label{sec5}

In this paper we have studied the roughness of a planar crack front driven
in a disordered medium, by considering the standard crack line model, with 
the additional ingredient of using toughness fluctuations with a 
characteristic scale $\xi$ which may be tuned. Such a simple approach has
made it possible to identify the relevant length scales of the problem,
and to propose an explanation of the experimental results regarding the
crack front roughness.

Based on our results, we think that the experiments reported in Ref. \cite{SAN-10}
correspond to the "strong pinning'' regime, where the toughness fluctuations
cause substantial local deformations of the crack front, visible as multiscaling
for short length scales.
We expect the cross-over scale separating this multiscaling regime from 
the asymptotic scaling regime characterized by the roughness exponent $\zeta
\approx 0.39$ to be proportional to the disorder scale $\xi$.
Indeed, in the experiments reported in Ref. \cite{SAN-10}, the cross-over scale
is shifted towards larger values when the diameter of the beads used for
the sand blasting process to prepare the experimental samples is increased.
Moreover, the spatial resolution of the experiment is clearly high enough
to observe features of the crack front also below scales corresponding to the
diameter of the beads \cite{SAN-10}. However, it is not clear how the 
statistical properties of the toughness fluctuations depend on the preparation 
procedure of the samples, making it difficult to make a precise comparison 
between theory and experiment.

Future directions of research related to the present topic would include
to develop an experimental setup in which the toughness fluctuations
could be controlled, and thus the effect of their statistical properties
could be tested properly. Regarding the theoretical side, it could be
useful to develop a model in which a more accurate expression for the long
range interaction kernel would be used, accurately describing interactions
between segments of a crack front of an arbitrary shape \cite{BOW-90}.
Such a more realistic line model for the crack front could also include
the possibility to form overhangs. However, we think that the picture
regarding the different scaling regimes and the cross-over scales separating
them presented in this paper would hold also in that case.

\ack
Stephane Santucci is thanked for interesting discussions. LL acknowledges
the financial support from the Academy of Finland.


\end{document}